\documentclass[12pt, letterpaper,twocolumn]{article} 
\usepackage{mathptmx}
\usepackage{geometry}
\usepackage{times}
\usepackage[utf8]{inputenc}
\usepackage{enumitem,amssymb}
\usepackage{multirow}
\usepackage{graphicx}
\usepackage{floatflt}
\newlist{thematic}{itemize}{8}
\setlist[thematic]{label=$\square$}
\usepackage{pifont}
\newcommand{\cmark}{\ding{51}}%
\newcommand{\done}{\rlap{$\square$}{\raisebox{2pt}{\large\hspace{1pt}\cmark}}%
\hspace{-2.5pt}}

\newcommand{\msun}{$M_{\odot}$}

\newcommand{\e}{et al.\ }
\newcommand{\etal}{et al.\ }

\newcommand{\skipthis}[1]{}

\newcommand\aj{{AJ}}
\newcommand\apj{{ApJ}}
\newcommand\apjl{{ApJL}}     
\newcommand\apjs{{ApJS}}
\newcommand\aap{{A\&A}}
\newcommand\mnras{{MNRAS}}

\usepackage{color}

\def\co2{\hbox{CO$_2$}}

\def\h2{\hbox{H$_2$}}

\begin{document}
\onecolumn
\raggedright
\huge
Astro2020 Science White Paper\\
\medskip
\LARGE
Understanding Galactic Star Formation with Next Generation X-ray Spectroscopy and Imaging 

\normalsize

\noindent \textbf{Thematic Areas:} \hspace*{60pt} $\done$ Planetary Systems \hspace*{10pt} $\done$ Star and Planet Formation \hspace*{20pt}\linebreak
$\square$ Formation and Evolution of Compact Objects \hspace*{31pt} $\square$ Cosmology and Fundamental Physics \linebreak
  $\done$  Stars and Stellar Evolution \hspace*{1pt} $\done$  Resolved Stellar Populations and their Environments \hspace*{40pt} \linebreak
  $\square$    Galaxy Evolution   \hspace*{45pt} $\square$             Multi-Messenger Astronomy and Astrophysics \hspace*{65pt} \linebreak
  
\textbf{Principal Author:}

Name: Scott J. Wolk	
 \linebreak						
Institution: Harvard--Smithsonian Center for Astrophysics  
 \linebreak
Email: swolk@cfa.harvard.edu 
 \linebreak
Phone:  617-496-7766
 \linebreak
 
\textbf{Co-authors:} 
Rachel Osten (STSCI, JHU),  Nancy Brickhouse (Harvard-Smithsonian Center for Astrophysics), Moritz G\"unther (Kavli Institute, MIT),  Laura A. Lopez (The Ohio State University), Jeremy Drake (Harvard-Smithsonian Center for Astrophysics), Benjamin F. Williams (University of Washington), Elaine Winston (Harvard-Smithsonian Center for Astrophysics), Denis Leahy (University of Calgary), Panayiotis Tzanavaris (NASA/GSFC-CRESST), David A. Principe (Kavli Institute, MIT)
\linebreak

\textbf{Abstract:}
This white paper is motivated by open questions in star formation, which can be uniquely addressed by high resolution X-ray imaging and require an X-ray observatory with large collecting area along good spectral resolution. A complete census of star-forming regions in X-rays combined with well matched infrared (IR) data 
will advance our understanding of disk survival times and dissipation mechanisms. In addition, we will be able to directly observe the effects of X-ray irradiation on circumstellar grain growth to compare with grain evolution models in both high- and low-UV environments. X-rays are native to stars at all phases of star formation and affect planet-forming disks especially through flares.  Moreover, X-rays trace magnetic fields which weave through the flares, providing a unique, non-gravitational feedback mechanism between disk and star.  Finally, the bright X-ray emission emanating from hot plasma associated with massive stars can have large scale impacts on the topology of star-forming regions and their interface with the interstellar medium (ISM). 
\twocolumn


\setcounter{page}{1}

\section*{Recent Progress in Star Formation} 

Understanding the fundamental aspects of star formation and evolution has been a major success of the \lq\lq Great Observatories\rq\rq\ era. Observations of star-forming regions are among the most challenging in modern astronomy. These regions are very crowded, with two-dimensional source densities exceeding 100 pc$^{-2}$, while their dynamic range is large, with pre-main sequence (PMS) brown dwarfs residing within an arcsecond of massive O stars. The overall range in stellar luminosity can easily cover 20 M$_V$. These regions are often filled with both hot gas -- which creates spectroscopic difficulties -- and cold dust -- which complicates the interpretation of the observed data and obscures some objects completely. \\
\medskip
Infrared (IR) missions such as {\it Herschel}, {\it Spitzer} and {\it 2MASS} have been critical in identifying young stellar objects (YSOs) with excess emission from envelopes and disks.  Modern sub-millimeter observatories, such as ALMA, can image some nearby disks in exquisite detail. Similarly JWST will reveal remarkable detail on specific systems. High resolution optical imaging and spectroscopy have already brought us new vistas of incredible systems such as HH30.   However, the optical data are limited to cases with fairly low dust obscuration 
and are mostly tracking scattered light. Meanwhile, the IR instruments only identify young stars while they are in possession of their natal disks, or in some cases, second generation debris disks. In contrast, X-ray indicators of stellar youth last for hundreds of  million to billions of years, covering a wide variety of masses and circumstances.  X-ray data serve not only as fingerprints of stellar youth, but the X-rays themselves are produced from the stellar dynamo and give important constraints on stellar magnetic field structure.

\begin{figure}
    \centering
    \includegraphics[scale=0.4]{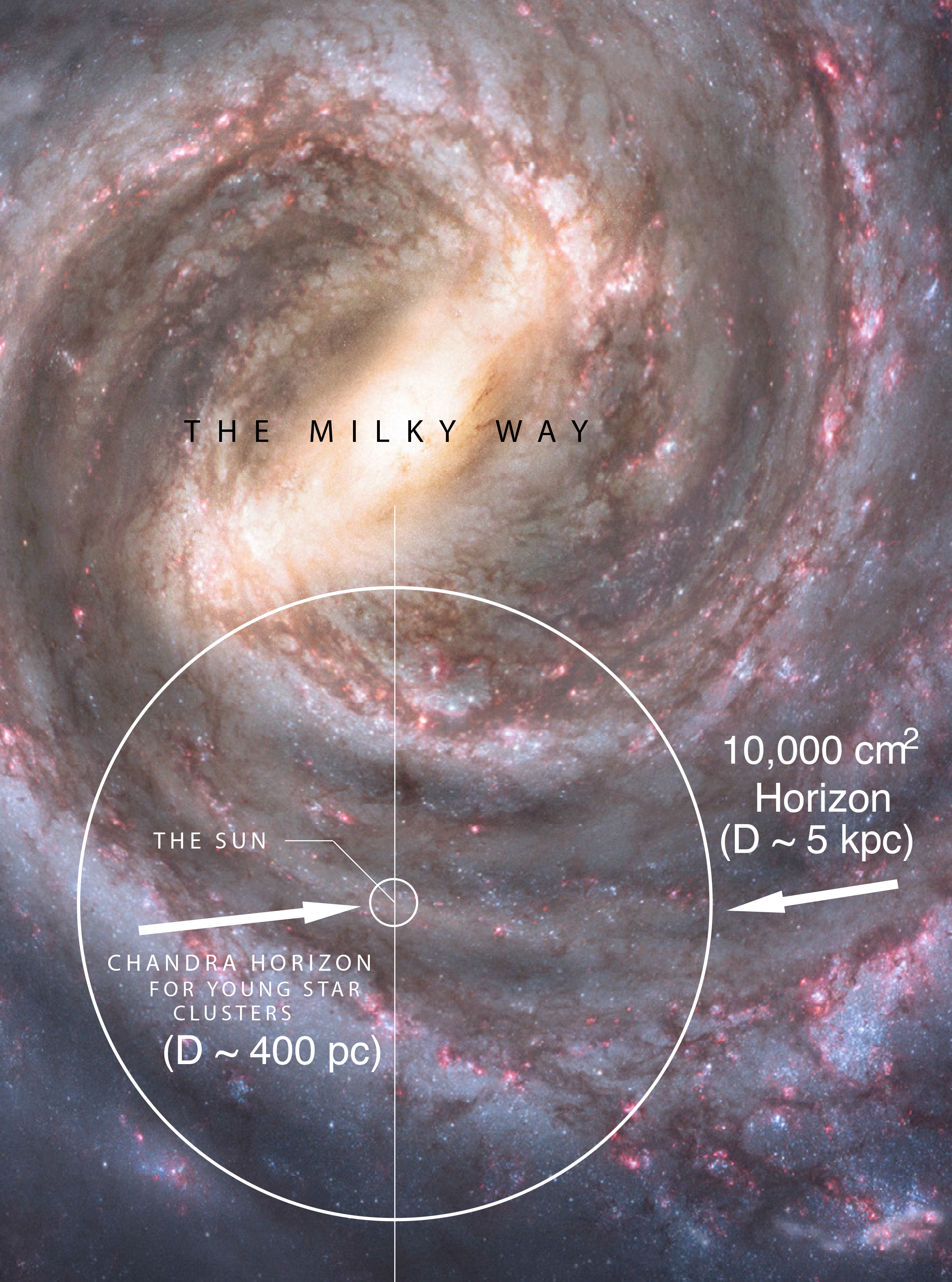}
    \caption{Graphic comparing the distance out to which {\it Chandra} ($\sim$ 600~cm$^2$ effective area) and a modern, 10,000~cm$^2$ effective area X-ray telescope are sensitive enough to take a complete census (from brown dwarfs to massive stars) of the young star clusters. While {\it Chandra} can only conduct a this inventory out to the Orion cluster ($D \approx$400~pc), an improvement in effective area with comparable spatial resolution would expand this reach to several kpc.}
    \label{fig:specres}
    \end{figure}
\vskip -1. in

\section*{The Role of X-Ray Studies} 

Young stars are bright, ubiquitous X-ray sources. For high-mass stars, most of the X-ray energy is coming from wind shocks and wind-wind collisions. The winds then interact with the ISM, depositing thermal energy and creating turbulence (e.g. MacLow \& Klessen 2004).  This turbulence inhibits other ongoing star formation.  For low-mass stars in the accretion phase, specific X-ray line ratios act as temperature and density diagnostics which allow us to directly infer the infall rate and constraining mechanisms (Brickhouse \e 2010).   This phase is also associated with enhanced flaring in which 100 MK long-lived flares are likely responsible for rapid heating of protoplanetary disks and can have deleterious effects on young planets.  Moreover, X-ray emission in general impacts circumstellar disk chemistry (Cleeves et al. 2015) and eventually disk dispersal (Owen et al. 2011). 

\medskip
For lower-mass stars, X-rays are generated in the corona which has its origin at the dynamo.  Modeling based on recent data indicates that the dynamo takes different forms depending upon the mass of the star (Yadav \e 2016).  For the most part, X-ray luminosity is seen to decrease with age following an exponential law with 
$L_X \propto t^{-0.5}$ 
(c.f. Favata \& Micela 2003).  This is driven by 
the convective turnover rate and indicates that late M stars have elevated X-ray fluxes for up to 5 billion years (Wright \e 2013).   
This long-term X-ray luminosity allows the identification of young stars long after they lose their dust disks.  Thus, stellar X-ray emission probes the evolution of star formation within existing clusters, and and allows us to study the star formation history of the whole Galaxy (Pillitteri \e 2013, Getman \e 2014).

\medskip
To date, Orion (at a distance of $\sim$400~pc) has been the main astrophysical laboratory where all the available observatories have been brought to bear. Studies of this cluster, and the relationship between its star-forming sub-regions and various triggering mechanisms -- from the youngest, deeply-embedded protostars discovered by {\it Herschel}, to the $\zeta$ Ori sub cluster, newly revealed in X-rays (Alves \& Buoy 2012, Megeath \e 2012, 2016, Getman \e 2005 etc.) -- have demonstrated the power of using well-matched observatories in concert. Multiwavelength facilities with comparable spatial resolution offer a complete view of the star-formation process and its relation to the ISM. In particular, X-rays are crucial to probe highly embedded sources, which have unique X-ray signatures. This is demonstrated by direct X-ray detections of protostars (c.f. Pravdo \etal 2009) as well as the detection of 6.4 KeV iron fluorescence, which is from circumstellar disk material being irradiated by stellar flares and exciting iron fluorescence (Tsujimoto \e 2005).  Consequently, it is possible to apply the reverberation mapping technique, used in studies of active galactic nuclei, to resolve AU-scale details on stellar disks that are several kpc from the sun.

\begin{figure}[htb]
\centering 
\includegraphics[width=0.45\textwidth]{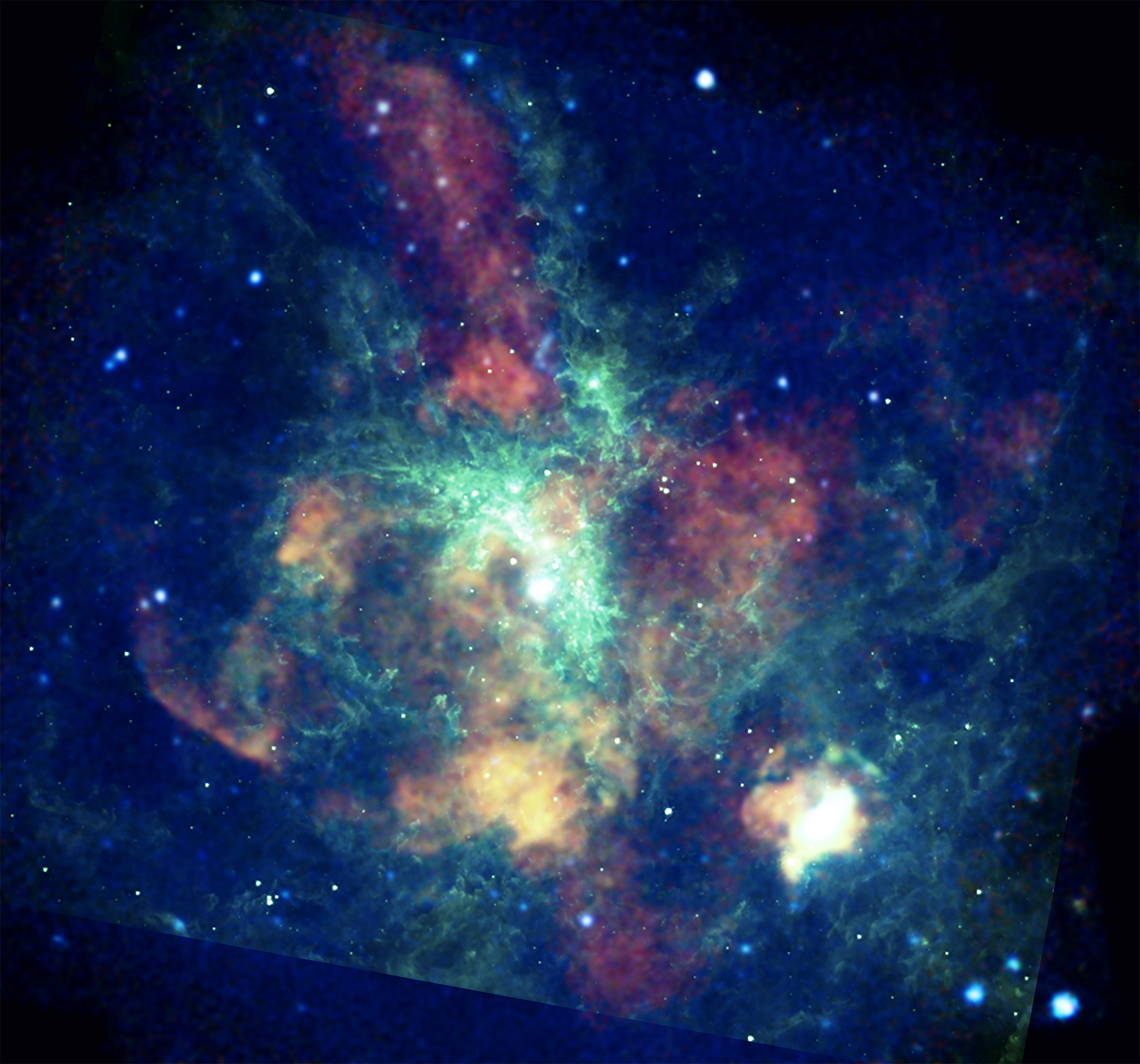}
\caption{(Left)
Composite image of 30 Doradus. Multi-million-degree gas detected in X-rays by~{\it Chandra} originating in shock fronts -- similar to sonic booms --formed by stellar winds and supernova explosions. In this rendering, 0.3-1.5 keV photons are red, 1.5-2.5 keV are green and 2.5-7 keV are blue. This hot gas carves out gigantic bubbles in the surrounding cooler gas and dust, shown here is IR dust emission from the {\it Spitzer} Space Telescope (green).
}
 \label{fig:30Dor}
\vspace{-0.0in}
\end{figure}

\medskip
A complete multi-wavelength and multi-cluster census will be much more impactful and informative than a study of a single cluster alone, no matter how well-studied.  For instance, data from Orion and the nearby clusters have been used to constrain circumstellar disk survival times, a crucial factor to halt planet formation. Such a study has been carried out for a relatively small number of clusters within 1 kpc of the Sun (Mamajek 2009), but has not been carried out for the more distant and more massive clusters. These kinds of studies naturally lead into the study of transition disk timescales.  The term {\it transition disk} derives from the fact that such YSOs either lack an IR excess at near- and mid- IR wavelengths (e.g. Wolk \& Walter 1996) or have spectral energy distribution (SED) slopes in-between those of typical star-forming disks and pure photospheres (e.g. Allen \e 2004). X-ray and high-energy flux appears to have a negative effect on grain growth in clusters (c.f. Winston \e 2010). Meanwhile many of the structures observed in broadband IR and optical observations seem to have their origin in an X-ray bright hot plasma expanding out into the ISM and being constrained by existing dust (c.f. Townsley \e 2003 Wolk \e 2002; Figure~\ref{fig:30Dor}). 

\section*{X-ray Astronomy Beyond the Next Decade} 

\begin{figure*}[htb]
\centering 
\includegraphics[width=0.9\textwidth]{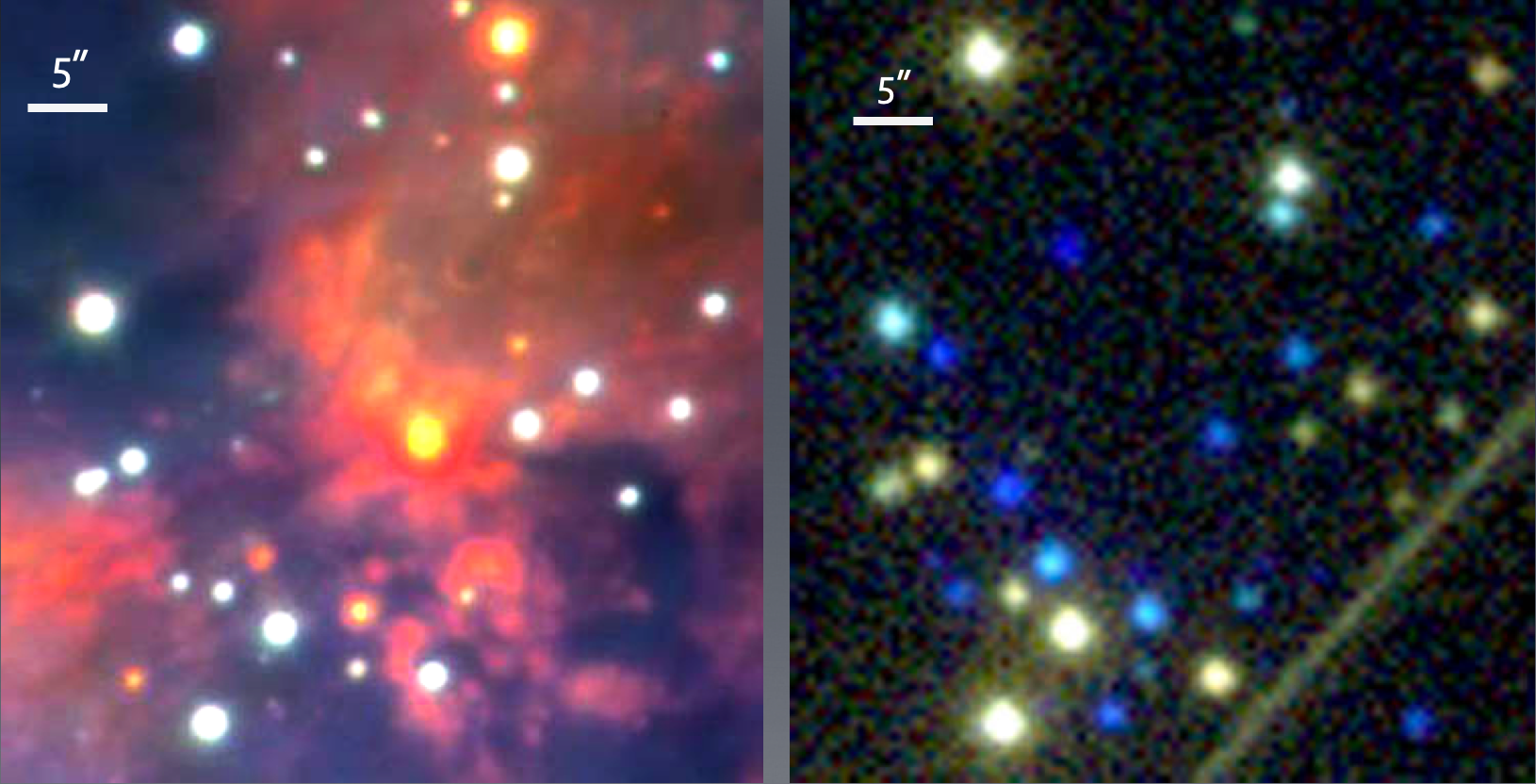}
\caption{
(Left) {\it VLT JKL} band image of the Becklin–Neugebauer Object in the Kleinmann–Low Nebula. Several deeply embedded objects
can be seen in red. (Right) The same region observed by {\it Chandra} with 0.5$^{\prime\prime}$ resolution.  The blue color is due to sources only detected in hard X-rays. This occurs partially because young stars have harder emission and softer X-rays are absorbed by gas near the star.  Note that hard X-ray sources track the red stars well but not perfectly.  Note the 5$^{\prime\prime}$ scale bar. This kind of analysis is only possible due to the good match between {\it Chandra}'s subarcsecond imaging and the resolution of {\it VLT} and most planned IR and optical missions in the 2020's.}
 \label{fig:BNKL}
\vspace{-0.0in}
\end{figure*}

While this white paper is directed at the 2020 Decadal committee, it is already likely that there will be no new major X-ray mission launched before $\sim$2030. Therefore the scientific problems to be solved will be already 10 years beyond today. By 2030, the current facilities {\it XMM-Netwon} and {\it Chandra} will be nearing end of life. While {\it e-ROSITA} will provide a deeper all-sky survey of X-ray sources than currently exists, it is not designed as a detailed followup instrument. {\it XRISM} will have limited sensitivity and spatial resolution, despite the groundbreaking spectral resolution expected from its microcalorimeter.  Athena should launch in the early 2030's but will lack the high spatial resolution need to understand compact regions such as the Kleinmann-Low Nebula even within 1 kpc (Figure~\ref{fig:BNKL}).  Arcsecond resolution is also required to maximize sensitivity.

\medskip
By 2030, many new missions will have made significant advances related to stars; however, each will have its own biases or limitations. {\it TESS} and {\it PLATO} will have surveyed millions of stars to find planets, and {\it JWST} and {\it WFIRST} will have provided deep IR and optical imaging and spectroscopy in crowded, star-forming fields down to well below the hydrogen burning limit. These observations will provide constraints based primarily on disk-bearing stars, leading to significant biases.  In the radio regime, {\it ALMA}, {\it SKA} and the next generation {\it VLA} will be making advances in the chemistry and structure of planet-forming disks. Large aperture ground based optical telescopes will be able to use Zeeman splitting to measure the surface magnetic fields in a variety of stars. {\it Gaia} may appear to be the most appropriate mission to identify YSOs in star-forming regions since, in theory, it will create a three-dimensional model of all the stars in every star-forming region within 10 kpc.  However, it should be noted that interstellar dust will greatly limit that range in the plane of the galaxy, where star formation is usually found.  And moreover, the local dust and ionized emission, which surround young stars, limits {\it Gaia}'s ability to map stars even in the closest clusters. 

\medskip

With the current generation of X-ray telescopes and those approved to be operating before 2030, Orion is the {\it only} massive cluster close enough to enable detailed study. However Orion is only one cluster.  It is relatively young, with an age of $\lesssim$20 Myr. As a single supercluster, it represents a single set of initial conditions.  Triggering mechanisms and total mass should affect the supernova rate and in turn the final elemental abundances and the energy budget returned to the ISM. It only represents a single sample of an IMF.  Even this is unclear as all the O stars seem to lie either in the Orion nebular cluster or in the belt.  On the other hand, there are thousands of young stars below the belt in Lynds 1641 without a proximate O or B star.      

In one simple example, our current understanding of disk survival times is limited by current X-ray sensitivity.  Kuhn \e (2014, 2015) studied archival {\it Spitzer} and {\it Chandra} data of about 15 regions of massive star formation. They find that IR-excess sources typically only represent about 50\%\ of the cluster. However they also find that the X-ray detection rates plummet with distance.  Most of the star-forming regions within the sensitivity limits of {\it Chandra} and {\it XMM-Newton} are within 2 kpc, the furthest being NGC 1893 at 3.6 kpc. This represents a selection bias since further clusters are simply ignored.  Worse still, the incompleteness begins to become a problem for X-ray sources below about 1~\msun\ at 2 kpc. A greater effective area will increase the grasp in the coming decade, but it will be important to also ensure the {\bf sub-arcsecond} imaging necessary to alleviate source confusion.  As the star-forming regions of interest move further away, the relative separations of stars shrink.  Moreover, as absolute stellar density increases with star cluster mass, spatial resolution will be crucial
with increasing X-ray telescope sensitivity (cf.\ Figure~\ref{fig:BNKL}).  \medskip

A high-spatial resolution, high collecting area X-ray telescope with a wide-field integral field spectrograph (IFS) will fundamentally change how we approach star formation in the X-rays and keep pace with expected improvements in technology already planned at other wavelengths.  Given an effective area of about 10,000 cm$^2$, an observation of Orion could reach the same depth as the 10-day (850ks) COUP (Getman \etal 2005) program in under one day of observing.  Assuming the data were taken with an IFS, each of the nearly 1000 stars detected in the central cluster region (perhaps 5$^{\prime}$ on a side) would have enough signal to measure not just one temperature but to detect {\it all} critical temperature- (and density-) sensitive lines such as Si~{\sc xiii}, Si~{\sc xiv}, Si~{\sc xv}, Mg~{\sc x}, Mg~{\sc xi}, Mg~{\sc xii}, O~{\sc vi}, O~{\sc vii}, O~{\sc viii}.  These data and the changes observed due to flares would be irreplaceable for regrading the high energy exposure of the inner disk and young (proto) exoplanets.  These conditions are fundamental to the local space environment and the formation and habitability of exoplanets. Further, given the greater sensitivity provided by both 10,000 cm$^2$ and arcsec imaging, Orion would no longer be the only star-forming region we could examine in such detail. Full surveys complete to one-tenth of a solar mass and including high quality multi-temperature and density sensitive spectra could be carried out on clusters, such as Carina (2.4 kpc), NGC 281 (2.1 kpc), NGC 3603 (6.9 kpc), Arches (8 kpc), and even 30 Doradus (50~kpc) in the Large Magellanic Cloud.  The duration of such a program would still be less than the amount of time spent by {\it Chandra} on Orion alone.

\section*{Summary} 
The next major X-ray facilities will not arrive until the 2030s. These facilities will join an era of star and planet formation science that will have been transformed by a fleet of missions and ground based facilities that will come of age in the 2020s. As a community, we cannot allow this high energy window into our Universe to close at a time when star and planet formation will be a major focus of the next generation of multiwavelength telescopes. X-rays will still play a fundamental role in our understanding of cluster morphology, stellar evolution, and the survival timescales of planet-forming disks and their dust evolution. Even the observed IMF will be out of reach without X-ray facilities working together with their multiwavelength, lower energy counterparts.  To pursue these science cases, the community needs major X-ray facilities that require effective area of $\approx$ 10,000 cm$^2$, an integral field spectrograph, and imaging quality on the order of an arcsecond.  


\small
\bibliographystyle{aasjournal}

\vskip 0.3in 
{\bf References} 

\noindent 
Allen, L.~E., et al.\ 2004, \apjs, 154, 363 

\vskip 0.1in\noindent 
Alves, J., \& Bouy, H.\ 2012, \aap, 547, A97

\vskip 0.1in\noindent 
Brickhouse, 
N.~S., Cranmer, S.~R., Dupree, A.~K., Luna, G.~J.~M., 
\& Wolk, S.\ 2010, \apj, 710, 1835

\vskip 0.1in\noindent 
Cleeves, L.~I., 
Bergin, E.~A., Qi, C., Adams, F.~C., 
\& {\"O}berg, K.~I.\ 2015, \apj, 799, 204 

\vskip 0.1in\noindent 
Favata, F., \& Micela, G.\ 2003, Space Science Reviews, 108, 577 

\vskip 0.1in\noindent 
Getman, K.~V., et al.\ 
2014, \apj, 787, 108 

\vskip 0.1in\noindent 
Getman, K.~V., 
Feigelson, E.~D., Grosso, N., McCaughrean, M.~J., Micela, G., Broos, P., 
Garmire, G., \& Townsley, L.\ 2005, \apjs, 160, 353 

\vskip 0.1in\noindent 
Kuhn, M.~A., Getman, K.~V., \& Feigelson, E.~D.\ 2015, \apj, 802, 60 

\vskip 0.1in\noindent 
Kuhn, M.~A., et al.\ 2014, 
\apj, 787, 107

\vskip 0.1in\noindent 
Mac Low, M.-M., \& Klessen, R.~S.\ 2004, Reviews of Modern Physics, 76, 125

\vskip 0.1in\noindent 
Mamajek, E.~E.\ 2009, American 
Institute of Physics Conference Series, 1158, 3

\vskip 0.1in\noindent 
Megeath, S.~T., et 
al.\ 2016, \aj, 151, 5 

\vskip 0.1in\noindent 
Megeath, S.~T., et 
al.\ 2012, \aj, 144, 192

\vskip 0.1in\noindent 
Owen, J.~E., Ercolano, B., \& Clarke, C.~J.\ 2011, \mnras, 412, 13 

\vskip 0.1in\noindent 
Pillitteri, I., 
et al.\ 2013, \apj, 768, 99 

\vskip 0.1in\noindent 
Pravdo, S.~H., Tsuboi, 
Y., Uzawa, A., \& Ezoe, Y.\ 2009, \apj, 704, 1495 

\vskip 0.1in\noindent 
Townsley, L.~K., 
Feigelson, E.~D., Montmerle, T., Broos, P.~S., Chu, Y.-H., 
\& Garmire, G.~P.\ 2003, \apj, 593, 874

\vskip 0.1in\noindent 
Tsujimoto, M., 
Feigelson, E.~D., Grosso, N., Micela, G., Tsuboi, Y., Favata, F., Shang, 
H., \& Kastner, J.~H.\ 2005, \apjs, 160, 503 

\vskip 0.1in\noindent 
Winston, E., et al.\ 
2010, \aj, 140, 266 

\vskip 0.1in\noindent 
Wolk, S.~J., Bourke, T.~L., 
Smith, R.~K., Spitzbart, B., \& Alves, J.\ 2002, \apjl, 580, L161

\vskip 0.1in\noindent
Wolk, S.~J., \& Walter, F.~M.\ 1996, \aj, 111, 2066

\vskip 0.1in\noindent
Wright, N.~J., Drake, 
J.~J., Mamajek, E.~E., 
\& Henry, G.~W.\ 2013, Astronomische Nachrichten, 334, 151

\vskip 0.1in\noindent
Yadav, R.~K., Christensen, 
U.~R., Wolk, S.~J., \& Poppenhaeger, K.\ 2016, \apjl, 833, L28



\end{document}